\documentclass[twocolumn,showpacs,preprintnumbers,amsmath,amssymb]{revtex4}

\usepackage{graphicx}
\usepackage{dcolumn}
\usepackage{bm}

\begin{document}

\preprint{APS/123-QED}

\title{Crossover of aging dynamics in polymer glass: from cumulative
aging to non-cumulative aging}

\author{K. Fukao}
 \email{kfukao@se.ritsumei.ac.jp}
 \altaffiliation[]{Present address: Department of
 Physics, Ritsumeikan University, Kusatsu, Shiga 525-8577 Japan}
\author{S. Yamawaki}%
\affiliation{%
Department of Polymer Science, Kyoto Institute of Technology, 
Matsugasaki, Kyoto 606-8585, Japan
}%


\date{\today}

\begin{abstract}
The aging behavior of polymer glass, poly(methyl methacrylate), has been
 investigated through the measurement of ac dielectric susceptibility at
 a fixed frequency after a temperature shift $\Delta T$ ($\le $ 20~K)
 between two temperatures, $T_1$ and $T_2$. A crossover from cumulative
 aging to non-cumulative aging could be observed with increasing $\Delta
 T$ using a twin temperature ($T$-) shift measurement. Based on a growth
 law of a dynamical coherent length given by activated dynamics, we
 obtained a unique coherent length for positive and negative $T$-shifts. The possibility of the existence of temperature chaos in polymer glasses is discussed.

\end{abstract}



\maketitle

\section{Introduction}
In amorphous materials, a glass transition
is observed when the temperature decreases from a high temperature 
in a liquid state to a lower temperature.
Below the glass transition temperature $T_g$, the structure of 
amorphous materials
is disordered, and the mobility due to a primary motion, which 
is usually called the $\alpha$-process, is lost. Thus, the
system is in a {\it glassy state}. However, even in the glassy state,
there is a very slow relaxation toward an equilibrium 
state~\cite{Struick,Bouchaud}. 
This slow relaxation is called aging and is regarded as an important 
common property characteristic of disordered materials, such as spin
glasses~\cite{Lefloch,Vincent,Jonason1,Jonason2}, orientational
glasses~\cite{Doussineau}, supercooled liquids~\cite{Leheny}, relaxor
ferroelectrics~\cite{Kircher}, and polymer
glasses~\cite{Bellon1,Bellon2}. 
Among the most interesting phenomena observed in the aging behavior
are {\it memory and rejuvenation} effects~\cite{Jonason2}.
The elucidation of these effects is essential for understanding 
the aging dynamics in glassy states and also the nature of the 
glass transition.

In our previous papers, we reported that the dielectric 
susceptibility of polymer glasses, poly(methyl methacrylate) (PMMA)~\cite{Fukao1}, 
and polystyrene (PS)~\cite{Fukao2} shows memory and rejuvenation
effects during an aging process below $T_g$. In these measurements, 
the system was quenched from a temperature above $T_g$ to an
initial temperature $T_1$, at which the system is aged for time $t_1$.
The temperature is then shifted to a lower temperature 
$T_2$ ($=T_1-\Delta T$) and maintained at this temperature for time $t_2$.
The temperature is then returned to and maintained at $T_1$.
Although in the first stage at $T_1$ the imaginary part of the
ac-dielectric susceptibility 
$\epsilon''(\omega;t_{\rm w})$ 
decreases with increasing aging time $t_{\rm w}$ 
(referred to as {\it aging}), 
it increases to a higher value and then begins to decrease again just after
the temperature is shifted to $T_2$. This behavior is called 
{\it rejuvenation}. If $\Delta T$ is large enough, 
the relaxation of dielectric susceptibility, which restarts at 
time $t_{\rm w}$=$t_1$, is approximately the same as that observed just after 
quenching directly from a high temperature ($>T_g$) to $T_2$.
Furthermore, at time $t_{\rm w}$=$t_1+t_2$, the value of the dielectric 
susceptibility returns to the value that $\epsilon''(\omega;t_{\rm w})$ had reached at time $t_{\rm w}$=$t_1$, and then begins to decrease as if the system 
had not experienced the lower temperature $T_2$. This is called
the {\it perfect memory effect}. 
The memory and rejuvenation effects are also observed in the real part
of ac-susceptibility $\epsilon'$ for PMMA.
It should be noted that the memory and 
rejuvenation effects on $\epsilon'$ and $\epsilon''$ observed in polymer
glasses are quite similar 
to those observed for ac-magnetic susceptibility
$\chi''(\omega;t_{\rm w})$
in spin glass systems, although there is a large difference in structure
between polymeric materials and spin glass. Therefore, it is thought that there is a universal physical phenomenon related to this aging behavior.

\section{Cumulative and non-cumulative aging}

An important concept for understanding glassy dynamics 
including memory and rejuvenation effects is the correlation length 
$\ell_T(t_{\rm w})$ after aging for time $t_{\rm w}$ at temperature $T$. 
This quantity was originally introduced to describe the 
spin-spin correlation in spin glass phase~\cite{Bouchaud,Berthier1}. 
In the present case, 
in which the dielectric susceptibility $\epsilon'(\omega;t_{\rm w})$ of
polymers is discussed, 
we can assume that the correlation length $\ell_T(t_{\rm w})$ describes 
the spatial correlation of dipole moments attached to the polymer
chains. In this case, the correlation length $\ell_T(t_{\rm w})$
suggests that there is an ordered domain with a characteristic 
size of $\ell_T(t_{\rm w})$, 
where spins or dipole moments are aligned in an orderly manner.

Here, we consider a temperature-shift ($T$-shift) protocol, in which
after the system is cooled down from a high temperature (above $T_{\rm g}$)
to $T_1$ (below $T_{\rm g}$) and is aged at $T_1$ for time $t_1$,
the temperature is rapidly shifted to $T_2$ and is maintained at $T_2$
for time $t_2$. 
This $T$-shift protocol is denoted as $(T_1,T_2)$.
During the $T$-shift protocol, we measure $\epsilon'(\omega;t)$.
We can then determine an effective time $t_{2,{\rm eff}}$ 
such that the relation
\begin{eqnarray}
\epsilon'(\omega;t_1+t_{\rm w})=\epsilon'_{T_2}(\omega;t_{2,{\rm
 eff}}+t_{\rm w})
\end{eqnarray}
is satisfied for any positive $t_{\rm w}$, 
where $\epsilon'_{T_2}$ is the dielectric 
susceptibility obtained 
for an isothermal aging process at $T_2$ for an aging time 
$t_{2,{\rm eff}}+t_{\rm w}$. In practice, $t_{2,{\rm eff}}$ is a fitting
parameter and is obtained from the observed data.
From this effective aging time, we can evaluate the effective correlation
length $\ell_{\rm eff}$ defined by the relation  
$\ell_{\rm eff}\equiv \ell_{T_2}(t_{2,{\rm eff}})$.
In a {\it cumulative aging} scenario, we can expect that
\begin{eqnarray}
\ell_{T_2}(t_{2,{\rm eff}})=\ell_{T_1}(t_1).
\end{eqnarray}
In this case, successive aging at the different temperatures 
$T_1$ and $T_2$ can be added in a totally cumulative manner,
although the growth rate of the correlation length may depend on 
the aging temperature.
Hence, the value of $t_{2,\rm eff}$ and $t_1$ can be described by
\[
 t_{2,\rm eff}=g(t_1,(T_1,T_2)),\quad t_1=g^{-1}(t_{2,{\rm eff}},(T_1,T_2))
\]
where $g$ is a function of $t_1$, which depends on the choice of
the initial and second temperatures, $T_1$ and $T_2$. Once a specific
form of $\ell_T(t)$ is given, the form of $g$ can be determined
uniquely. In this case, there is a reversible aging between two
temperatures. If this reversibility is violated, the aging is no longer
cumulative, but rather it is {\it non-cumulative}. 

Here, we introduce a reverse $T$-shift protocol ($T_2$, $T_1$), 
where the system is aged at $T_2$ for time $t_2$, and the 
temperature then is shifted from $T_2$ to $T_1$. 
For this protocol, we can obtain another effective
aging time $t_{1,{\rm eff}}$, which satisfies the relation
\begin{eqnarray}
\epsilon'(\omega;t_2+t_{\rm w})=\epsilon'_{T_1}(\omega;t_{1,{\rm
 eff}}+t_{\rm w}).
\end{eqnarray}
In the case of cumulative aging, we can also expect that
\begin{eqnarray}
\ell_{T_1}(t_{1,{\rm eff}})=\ell_{T_2}(t_2).
\end{eqnarray}

In order to check whether the aging is cumulative or non-cumulative,
{\it a twin $T$-shift experiment} has been proposed in the
studies on spin glass by J\"onsson, Yoshino, and Nordbald~\cite{Jonsson2}.
In the twin $T$-shift experiment, a pair of $T$-shift experiments 
($T_1$, $T_2$) and ($T_2$, $T_1$) are performed. 
Then, an effective aging time $t_{2,{\rm eff}}$ ($t_{1,{\rm eff}}$) 
is determined as a function of the aging time $t_1$ ($t_2$) at 
$T_1$ ($T_2$) for the $T$-shift ($T_1$,$T_2$) (($T_2$,$T_1$)),
and the data points ($t_1$,$t_{2,{\rm eff}}$) and 
($t_{1,{\rm eff}}$,$t_2$) for various values of $t_1$ and $t_2$ 
are plotted on the same plot, where
the horizontal axis is $t_1$ or $t_{1,{\rm eff}}$ and the vertical
axis is $t_{2,{\rm eff}}$ or $t_2$, as shown in Fig.~4. If all data points 
fall onto a single curve, we can judge that the aging is cumulative.
Otherwise, the aging is non-cumulative.

In the present study, we have performed twin $T$-shift experiments on 
the ac-dielectric susceptibilities during the aging process of PMMA glasses
in order to check whether the aging dynamics are attributed to
cumulative or non-cumulative aging and also to check whether there is
a crossover between these types.
The present paper consists of six sections.
After providing an explanation on cumulative aging in Section II and 
experimental details in Section III, 
experimental results on twin $T$-shift measurements 
are presented in Sections IV and V. Finally,
a discussion on the experimental results is presented 
in Section VI.

\section{Experiment}
Polymer samples used in the present study are atactic
PMMA purchased from Scientific Polymer Products, Inc. 
The weight-averaged molecular
weight is $M_{\rm w}$~=~3.56$\times$10$^5$, and $M_{\rm w}/M_{\rm n}$~=~1.07,
where $M_{\rm n}$ is the number-averaged molecular weight.
The glass transition temperature $T_g$ determined by 
differential scanning calorimetry (DSC) is approximately 380~K.
Thin films were prepared onto an aluminum vacuum-deposited glass
substrate using the spin coat method from a toluene solution of PMMA.
After annealing at 343 K, aluminum was vacuum deposited
again to serve as an upper electrode. 
The thickness of the films used in the present study 
is 281~nm, as measured directly by atomic force microscopy.
The preparation method of the above samples is the same
as in our previous studies~\cite{Fukao3,Fukao4,Fukao5}.

Dielectric measurements were performed using 
an LCR meter (HP4284A). The frequency range of the applied electric field was
from 20~Hz to 1~MHz, and the applied voltage level was 2.0 V.
In our measurements, the complex electric capacitance of the sample
condenser $C^*$ was measured and then converted into 
the ac-dielectric susceptibility $\epsilon^*$ by dividing $C^*$ by the 
geometrical capacitance $C_0$ at a standard temperature $T_0$.
Here, $C^*$ is given by $C^*=\epsilon^*\epsilon_0\frac{S}{d}$ and
$C_0=\epsilon_0 \frac{S}{d}$, 
where $\epsilon_0$ is the permittivity in vacuum, $S$ is the area of 
the electrode, and $d$ 
is the film thickness. For the evaluation of $\epsilon^*$ and $C_0$, 
we use the thickness $d$ given above and 
$S$~=~8$\times$10$^{-6}$~m$^2$.
The sample prepared above is located in a sample cell. 
The temperature of the sample cell is controlled through
heaters wound along the outer surface of the sample cell.
The temperature measured by a thermo-couple attached to the backside
of glass substrate is used as the temperature of the sample.

Heating cycles in which the temperature was varied between 
room temperature and 403~K ($>T_{\rm g}$) were applied 
several times 
prior to the measurements, in order to relax the as-prepared samples and 
obtain reproducible results.
The relaxed sample is heated from room temperature to 403~K
and then cooled to an aging temperature $T_1$
($T_2$) at the rate of 0.5~K/min. The temperature is thereafter maintained 
at $T_1$ ($T_2$) for a time $t_1$ ($t_2$). The temperature is then changed to $T_2$ ($T_1$) at the rate of 0.5~K/min and is then maintained at $T_2$ ($T_1$) for
30~h. Here, the aging times $t_1$ and $t_2$ change from 1~h to 20~h.
For the twin $T$-shift measurements, during this thermal history, 
dielectric measurements between 1~MHz and 20~Hz are performed
repeatedly. One measurement for this frequency range requires approximately 50~sec.

\section{Determination of effective aging times}

Figure 1 shows an example of the results observed for a positive 
$T$-shift from $T_2$~=~356.9~K to $T_1$~=~375.7~K
after aging at $T_2$ for a time $t_2$~=~10~h.
The vertical axis is the real part of ac-dielectric susceptibility
$\epsilon'$ relative to a reference value $\epsilon'_{\rm ref}$. 
The reference value for $T_1$ ($T_2$) is the value measured 
at $T_1$ ($T_2$) after direct quenching from 
403~K to $T_1$ ($T_2$). 
Both values are measured at a fixed frequency of 100 Hz.
Figure 1(a) shows that after the temperature 
reaches $T_2$
from 403~K, a decrease in $\epsilon'$ from the reference 
value $\epsilon'_{\rm ref}$ occurs, 
and as the temperature changes from $T_2$ to $T_1$
($T_1>T_2$) at time $t_2$ the value of $\epsilon'$ approaches 
the reference value
$\epsilon'_{\rm ref}$ and then begins to decrease with increasing 
aging time after reaching the maximum. This relaxation behavior 
deviates from that observed for isothermal aging at $T_2$.

\begin{figure}
\includegraphics[width=8.5cm,angle=0]{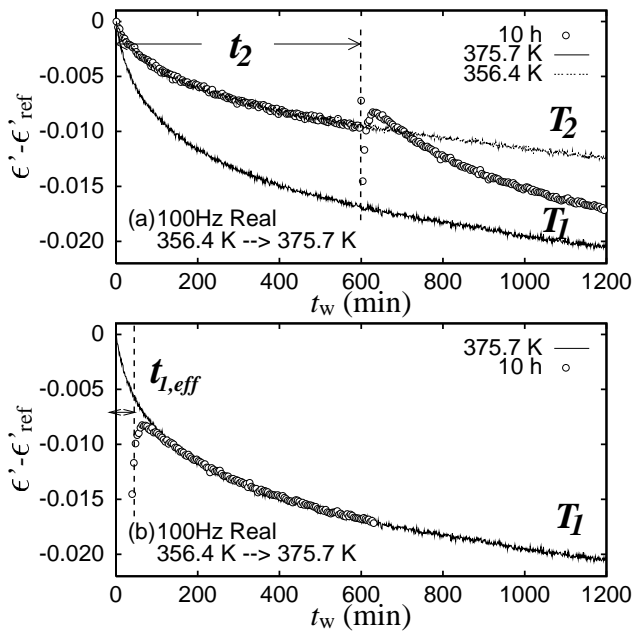}
\caption{\label{fig:1}(a) Difference between $\epsilon'$ and
 $\epsilon'_{\rm ref}$ observed for the positive $T$-shift
 measurements. $T_1$=~375.7~K, and $T_2$ =~356.4~K ($\Delta
 T$~=~19.3~K). The data are obtained at 100~Hz. (b)
 $\epsilon'-\epsilon'_{\rm ref}$ obtained by shifting the data points
 after the $T$-shift in the negative direction of the time axis by $\tau_1$,
 so that the data points after the $T$-shift are well-overlapped with
 the standard relaxation curve at $T_1$.}
\end{figure}

In order to compare the change in $\epsilon'$ with the aging time for
$t_{\rm w}>t_2$, 
the data points for $t_{\rm w}<t_2$ are removed and those for 
$t_{\rm w}>t_2$ are
shifted to the negative direction of the $t_{\rm w}$-axis 
by a certain amount $\tau_1$ so that the data points for $t_{\rm w}>t_2$ 
can be well-overlapped with those observed for the isothermal aging process
after directly quenching from 403~K to $T_1$, except for a small time
region just after $t_{\rm w}=t_2$. 
In this case, an effective time $t_{1,{\rm eff}}$ can be obtained 
by the following relation:
\[
 t_{1,{\rm eff}}=t_2-\tau_1.
\]
For the examples shown in Fig.~1, it is found that 
$t_{1,{\rm eff}}$ = 0.6~h, which means that the aging at $T_2$ = 356.4~K 
for a time $t_2$ = 10~h 
corresponds to the aging at $T_1$ = 375.7~K for a time 
$t_{1,{\rm eff}}$ = 0.6~h. Figure~1(b) shows that the data points for
$t_{\rm w}>t_2$
can be well-overlapped with the curve for $T_1$ after shifting by
$\tau_1$, except for a small time region just after the temperature shift.

\begin{figure}
\includegraphics[width=8.5cm,angle=0]{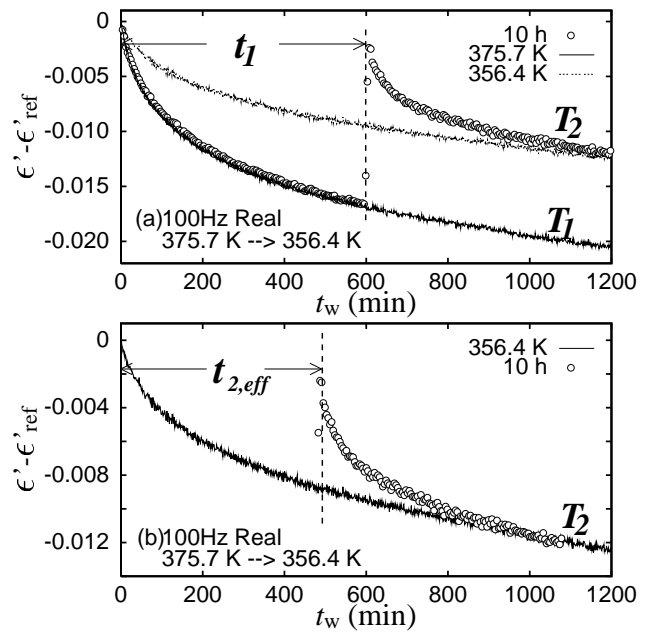}
\caption{\label{fig:2} (a) Difference between $\epsilon'$ and
 $\epsilon'_{\rm ref}$ observed for the negative $T$-shift measurement
 from 375.7~K to 356.4~K. (b) $\epsilon'-\epsilon'_{\rm ref}$ obtained
 by shifting the data points after the $T$-shift in the negative direction
 of the time axis by $\tau_2$, in the same manner, as shown in Fig.~1(b).
}
\end{figure}

For a negative $T$-shift from $T_1$ = 375.7~K to $T_2$ = 356.4~K,
we can evaluate another effective temperature $t_{2,{\rm eff}}$ 
in a manner similar to that for the positive $T$-shift, as shown in Fig.~2(b).
Here, the value of $\epsilon'$, which is shifted to the negative direction of 
the $t_{\rm w}$-axis by $\tau_2$. 
In this case, for a certain value of $\tau_2$, the data points at 
$t_{\rm w}>t_1$ merge to the reference curve for the value of $T_2$ obtained above,
and thereafter lie on this curve. We can determine an effective time 
$t_{2,{\rm eff}}$ as follows: $t_{2,{\rm eff}}=t_1-\tau_2$.
Figure 2 indicates that $t_{2,{\rm eff}}$ = 8~h, which means that the aging at 
$T_1$ = 375.7~K for a time $t_1$ = 10~h corresponds to the aging at $T_2$ = 356.4~K for a time $t_{2,{\rm eff}}$ = 8~h. Figure 2(b) shows that the data points for
$t_{\rm w}>t_1$ can be well-overlapped with the curve for $T_2$ after shifting by
$\tau_2$, except for a time region ($t_{2,{\rm eff}}<t_{\rm w}< 2t_{2,{\rm eff}}$) 
after the temperature shift.
The deviation of $\epsilon'-\epsilon'_{\rm ref}$ from the reference curve at
$T_2$ for the time range just after $t_{2,{\rm eff}}$ in Fig.~2(b) will be compared with that obtained in spin glass and discussed later herein.

\section{twin-$T$ shift experiments}

\begin{figure}
\includegraphics[width=8.5cm,angle=0]{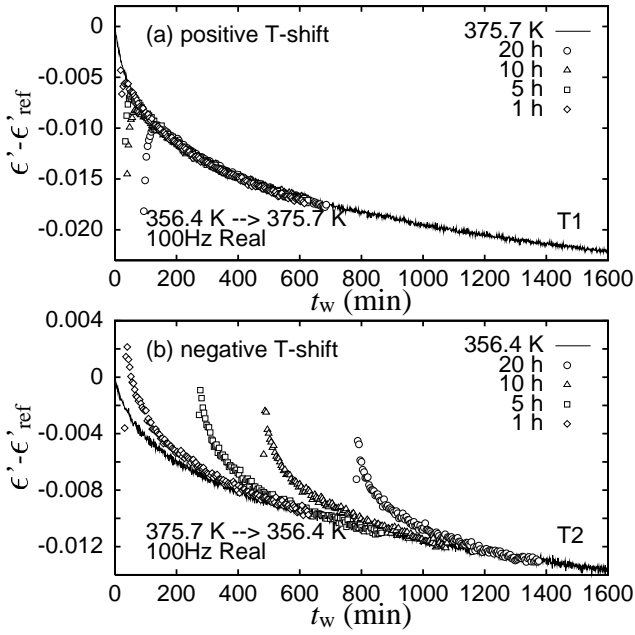}
\caption{\label{fig:3} (a) Values of $\epsilon'-\epsilon'_{\rm ref}$
 obtained by shifting the data points by a certain amount of time in
 order to obtain the effective aging times $t_{1,{\rm eff}}$ (Fig.~3(a))
 and $t_{2,{\rm eff}}$ (Fig.~3(b)) for various aging times before the
 $T$-shift. $\Delta T$ = 19.3~K.}
\end{figure}

Figure 3 shows the real part of the ac-dielectric susceptibility at
100~Hz as a function of aging
time after positive and negative temperature shifts with various values
of $t_1$ ($t_2$) from 20~h to 1~h and $\Delta T$ =~19.3~K. 
The solid curves show the reference curve observed at $T_1$ ($T_2$)
after quenching directly from high temperature to the aging temperature.
The data points are shifted in the
negative direction of the $t_{\rm w}$-axis by a certain amount so that they will fall on top of the reference curve for $T_1$ ($T_2$).
It is found that the effective aging times $t_{1,{\rm eff}}$ 
($t_{2,{\rm eff}}$) increase with increasing aging times $t_2$ ($t_1$)
at $T_2$ ($T_1$). From these results, we can extract the effective times as a function of aging time.

\begin{figure}
\vspace*{-2.2cm}
\hspace*{-1.5cm}
\vspace*{-2cm}
\includegraphics[width=8.5cm,angle=0]{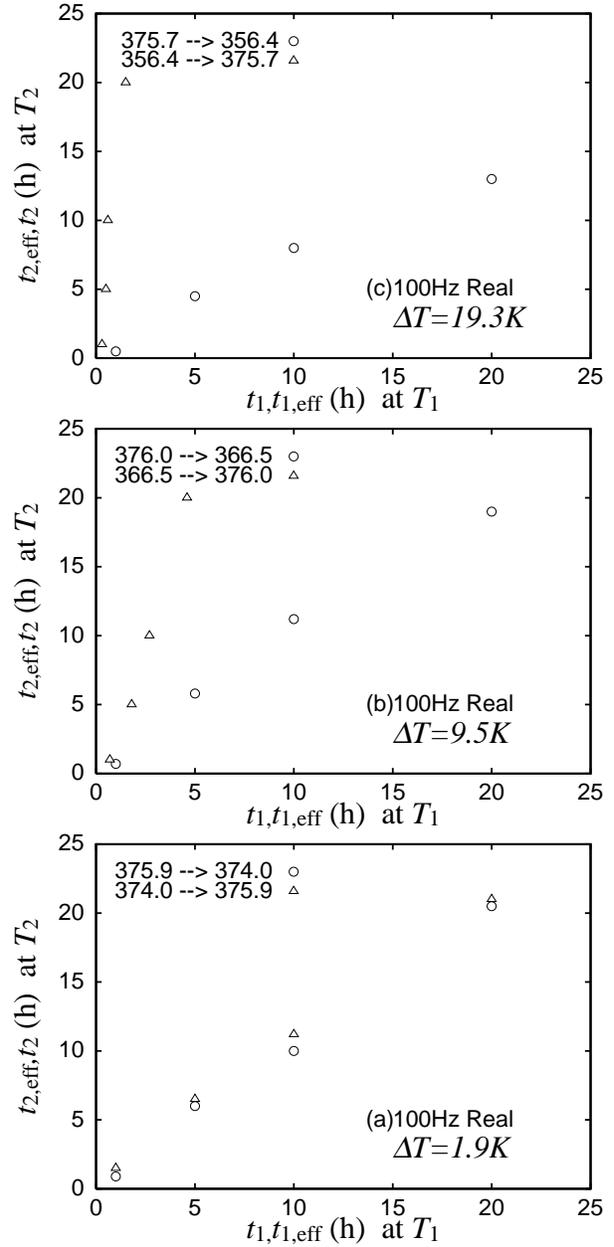}
\caption{\label{fig:4} Relation between $t_{\rm w}$ and $t_{\rm eff}$ in twin $T$-shift experiments ($T_1$, $T_2$) and ($T_2$, $T_1$). $T_1$~=~375.7~K, and $T_2$~=~$T_1-\Delta T$, where $\Delta T$~=~(a)~1.9~K, (b)~9.5~K, and (c)~19.3~K. This figure shows that there is a crossover from cumulative aging to non-cumulative aging with increasing $\Delta T$. The data in this figure are evaluated based on the real part of the dielectric constant at 100~Hz.}
\end{figure}

From a pair of negative and positive $T$-shift measurements, we obtain
two effective times $t_{1,{\rm eff}}$ and $t_{2,{\rm eff}}$ as 
a function of aging times $t_2$ and $t_1$, respectively.
In Fig.~4 the horizontal axis is the time $t_1$ or 
$t_{1,{\rm eff}}$ for $T_1$, and the vertical axis is the time 
$t_{2,{\rm eff}}$ or $t_2$ for $T_2$. The data points 
($t_1$,$t_{2,{\rm eff}}$) and ($t_{1,{\rm eff}}$,$t_2$) are plotted
in Fig.~4 for various values of $\Delta T\equiv |T_1-T_2|$.
In Fig.~4(a), the data points for the negative $T$-shift
fall on the same line as do the data points for the positive $T$-shift
for $\Delta T$= 1.9~K. This suggests that the aging dynamics for 
$\Delta T$ = 1.9~K obey {\it a cumulative aging}, 
where the contribution from the aging at $T_1$ to the coherent dynamic length scale and that from the aging at $T_2$ are totally additive, 
although the rate of aging dynamics differs depending on
temperature. If $\Delta T$ increases from 1.9~K to 9.5~K and 19.3~K,
it is clear that the data points obtained for the positive temperature
shift do not fall on the same line as those for the negative
temperature shift. This suggests that there is no reversibility 
between the positive and negative temperature shift, and hence,
the aging dynamics for $\Delta T$ = 9.5~K and 19.3~K can be
regarded as non-cumulative aging. Therefore, there is {\it a crossover from the cumulative aging to the non-cumulative aging} with increasing $\Delta T$ for aging dynamics in glassy states in PMMA.

\section{Discussions} 

In order to further analyze the present data,
a growth law of dynamical coherent length $\ell_{T}(t)$ is required.
This coherent length can be regarded as the linear size of the 
domain in which dipole moments are aligned in an orderly manner.
Here, we assume that the growth of this ordered domain 
proceeds if the system exceeds a potential barrier $E(\ell)$,
which is a function of the dynamic coherent length 
$\ell\equiv\ell_{T}(t)$, and that the following Arrhenius relation holds:
\[
 t=t_0\exp\left(\frac{E(\ell)}{k_BT}\right),
\]
where $t$ is the elapsing time at temperature $T$, $t_0$ is a microscopic
time scale, and $k_B$ is the Boltzmann constant~\cite{Bouchaud}. 
If the potential energy barrier is given by 
\[
 E(\ell)=E_0\left(\frac{\ell}{\ell_0}\right)^{\psi},
\]
where $E_0$ is a constant factor, $\ell_0$ is the unit length, 
and $\psi$ is a constant. We can obtain the following growth law
of the ordered domain:
\[
 \ell_{T}(t)=\ell_0\left(\frac{k_BT}{E_0}\log\left(\frac{t}{t_0}\right)\right)^{1/\psi}.
\]
Here, the scale dependence of the potential energy barrier can 
often be seen in problems of elastic objects, such as domain walls 
pinned by random impurities~\cite{Bouchaud}.
The constant factor $E_0$ can be dependent on temperature.

Next, we choose the parameters as follows:  
unit length $\ell_0$=1, exponent $\psi$=1, and temperature unit $T_0\equiv E_0/k_B$=$T_g$ (in this case, $T_g$=380 K for PMMA).
Using this growth law of the coherent length,
we can convert both the aging time $t_i$ and 
the corresponding effective time $t_{j,{\rm eff}}$ 
into the domain sizes $\ell_{T_i}(t_i)$ or 
$\ell_{T_j}(t_{j,{\rm eff}})$ ($i$, $j$= 1 or 2).
For the positive $T$-shift, we can obtain
($\ell_{T_1}(t_1)$,$\ell_{T_2}(t_{2,{\rm eff}})$) from 
($t_1$,$t_{2,{\rm eff}}$), and for the negative $T$-shift, 
we can obtain ($\ell_{T_2}(t_2)$,$\ell_{T_1}(t_{1,{\rm eff}})$) from 
($t_2$,$t_{1,{\rm eff}}$).

\begin{figure}
\vspace*{-0.1cm}
\hspace*{-1.5cm}
\includegraphics[width=8.5cm,angle=0]{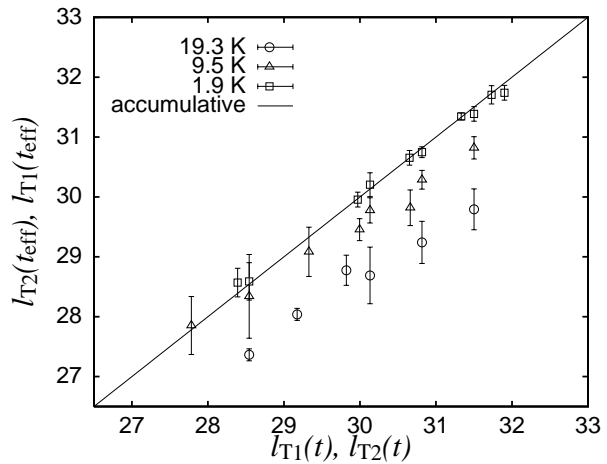}
\caption{\label{fig:5} Relationship between coherent length and
 effective coherent length obtained in the twin $T$-shift experiments for
 various values of $\Delta T$ (1.9~K, 9.5~K, and 19.3~K). The straight
 line indicates the case of cumulative aging. The error bars show
 the extent of data points for ac-electric fields of various frequencies 
(100~Hz, 1~kHz, and 10~kHz) and also for the choice of the real or
 imaginary parts of complex dielectric constant.
}
\end{figure}

Figure 5 shows the relation between $\ell_{T_i}(t_i)$ and
$\ell_{T_j}(t_{j,{\rm eff}})$ for the twin $T$-shift experiments for
three different values of $\Delta T$ ($i$, $j$ = 1 or 2). 
The data points obtained from both the
real and imaginary parts of the dielectric constant for three different
frequencies of the applied electric field, 100~Hz, 1~kHz, and 10~kHz,
are included.
Error bars attached to the data points represent the variation for
various conditions.
The straight line corresponds to the case of fully cumulative
aging. The results for $\Delta T$= 1.9~K clearly shows that the observed
values fall on the line corresponding to the cumulative aging. 
However, as the values of $\Delta T$ increase from 1.9~K, the data 
points shown in Fig.~5 deviate from the straight line corresponding to the
cumulative aging.
Note also that, for a fixed value of $\Delta T$, the data points
fall on the same curve within error bars independent of whether the
temperature
shift is negative or positive, or of which frequency is used for the
capacitance measurements. This indicates that there is a unique 
coherent length for both the negative and positive $T$-shift
measurements of the aging process.

Within the framework of the droplet theory, there is an overlap length
$\ell_{\Delta T}$ for a given 
$\Delta T$($\equiv|T_1-T_2|$)~\cite{Bray1,Fisher1,Fisher2,Fisher3}.
If the temperature-chaos effect exists, one can expect that the aging is
totally uncorrelated between the two temperatures $T_1$ and $T_2$
for a length scale beyond $\ell_{\Delta T}$, while the aging 
can be regarded as cumulative for a length scale far below
$\ell_{\Delta T}$.  In spin glasses, the existence of the overlap length is one possible origin for the existence of the rejuvenation phenomenon~\cite{Jonsson2}.
In this case, we can expect that $\ell_{\rm
eff}(\equiv \ell_{T_j}(t_{j,{\rm eff}}))$ 
can be scaled as follows:
\[
 \frac{\ell_{\rm eff}}{\ell_{\Delta
 T}}=f\left(\frac{\ell_{T_i}(t_{i,{\rm eff}})}{\ell_{\Delta T}}\right),
\]
where a scaling function $f(x)$ satisfies the relation $f(x)=x$ for
$x << 1$ and $f(x)=1$ for $x >> 1$ and $i$, $j$ = 1 or 2~\cite{Jonsson2}.
This scaling function suggests that the coherent length cannot grow 
beyond the overlap length $\ell_{\Delta T}$, but it merges to
$\ell_{\Delta T}$. In the present case, we assume that there is an 
overlap length scale $\ell_{\Delta T}\approx 
\left|\frac{T_0}{\Delta T}\right|^{\gamma}$ 
with the parameters $\ell_0=1$, $\gamma=0.2$, and $T_0=T_g=380 K$ for 
PMMA. Then, we can obtain the scaling behavior of $\ell_{T_j}(t_{j,{\rm eff}})$, as shown in Fig.~6. This scaling behavior is originally discussed for spin glass, and this scaling has been regarded as experimental evidence for the existence
of the temperature chaos effect.

\begin{figure}
\vspace*{-0.1cm}
\hspace*{-1.5cm}
\includegraphics[width=8.5cm,angle=0]{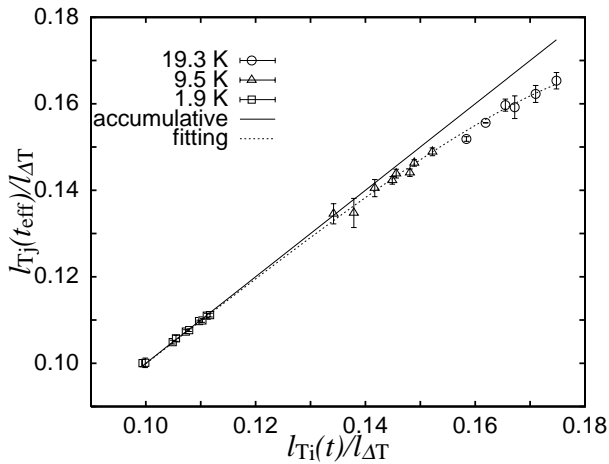}
\caption{\label{fig:6} Scaling behavior of the coherent length scale $\ell_T(t)$. Here, the overlap length is given by $\ell_{\Delta T}=\tilde\ell | \frac{T_0}{\Delta T}|^{\gamma}$, where $\tilde\ell$=100. The solid line corresponds to the case of cumulative aging. The dashed curve is for obtained data fitted with a model function.}
\end{figure}

The scaling behavior in Fig.~6, for the case of PMMA glass, suggests that
the same argument based on the temperature chaos can be used for the 
existence of the rejuvenation phenomena, even in polymer glasses. This is 
very surprising if we take into account the large difference in
structure between polymer glasses and spin glass. In the literature, it
has been reported that 
a temperature chaos can exist in polymer chains ~\cite{Yoshino}.

As shown in Figs.~1 and 2, after the positive $T$-shift, the shifted 
dielectric susceptibility approaches the standard relaxation curve at 
$T_1$ {\it from the lower side}, whereas after the negative $T$-shift, the shifted 
dielectric susceptibility approaches the standard relaxation curve at $T_2$ {\it from the upper side}. This behavior is observed in the non-equilibrium spin-glass dynamics of a strongly interacting ferromagnetic nanoparticle system (superspin
glass), in which no evidence of temperature chaos or strong rejuvenation has been reported. Although the present results suggest the possibility of the existence of temperature chaos in polymer glasses, the above-described differences between spin glasses and polymer glasses should also be resolved in the near future.

Finally, the microscopic origin of the decrease in dielectric susceptibility 
should be discussed. In the case of PMMA, the dielectric susceptibility
$\epsilon'$ (and $\epsilon''$) decreases with aging time during the 
isothermal aging process~\cite{Bellon1,Fukao1}. 
This decrease can be observed also in other polymeric systems
such as polycarbonate~\cite{Cangialosi} and poly(ethylene
terephthalate)~\cite{McGonigle}.  Hence, the decrease in the dielectric
susceptibility during isothermal aging process is a characteristic property 
common for physical aging of various polymeric systems, 
although the microscopic origin of this decrease is not fully understood. 
 
It is well-known that for many polymeric systems the densification
occurrs due to physical aging. Hence, the density should increase
during aging process and accordingly the dielectric strength 
$\Delta\epsilon$ should increase because the $\Delta\epsilon$. 
Several dielectric measurements under high pressure for 
polymeric systems show an increase in dielectric constat upon 
high pressure~\cite{Floudas1,Floudas2,Floudas3}.
In this case, it is expected that the densification occurs, and hence 
the dielectric susceptibilities under high pressure might be compared to those
during aging process. However, the results under high pressure are opposite to
those in our measurements. Therefore, there must be another physical
origin which decreases dielectric susceptibility and overcomes the
densification.

During the course of physical aging two possible
scenarios are possible: 1) physical aging is probed well below
the glass transition temperature, implying that the $\alpha$-process 
is totally frozen and only secondary relaxation processes, 
associated with localized motion, are dynamically active. 
In this case, orientational polarization associated
with these processes would be reduced during physical aging; 2)
physical aging is probed just below the glass transition temperature 
and the high frequency tail of the $\alpha$-process is probed.  
In this latter case the decrease in the dielectric susceptibility is 
simply due to a shift of this tail to lower
frequency~\cite{Lunkenheimer}. 
In the case of PMMA, there is a contribution from the $\beta$-process
below $T_g$ which is associateed with a localized motion of 
branches attached to polymer chains. Furthermore, measurements in the
frequency domain of the dielecric loss clearly shows that the peak 
frequency of the $\beta$-process is not shifted, but the peak height 
decreases with increasing aging time without changing the shape of
the dielectric spectrum during the isothermal aging process. 
The microscopic origin for the decrease in dielectric susceptibility  
in this case might be mainly attributed to the $\beta$-process, not to the
$\alpha$-process.

\acknowledgments
The authors would like to appreciate H.~Yoshino and H.~Takayama for useful
discussions.
This work was supported by a Grant-in-Aid for Scientific Research
(B) (No. 16340122) from the Japan Society for the Promotion of Science,
and also by KAKENHI (Grant-in-Aid for Scientific Research) on 
Priority Area ``Soft Matter Physics'' from the Ministry of Education, 
Culture, Sports, Science and Technology of Japan.
\\




\end{document}